\documentclass[11pt]{article}
\pdfoutput=1
\usepackage[breaklinks=true]{hyperref}
\usepackage{graphicx}
\usepackage{epsfig}
\usepackage{dcolumn}
\usepackage{subfigure}
\usepackage{amsmath,amsthm, amssymb}
\usepackage{slashed}

\usepackage{comment}

\newcommand{\bea}{\begin{eqnarray}}
\newcommand{\eea}{\end{eqnarray}}
\newcommand{\beq}{\begin{equation}}
\newcommand{\eeq}{\end{equation}}
\newcommand{\nn}{\nonumber}

\newcommand{\BB}[1]{{\mathbb{#1}}}

\newcommand{\braket}[2]{\langle #1\vert\,#2\rangle}
\newcommand{\avg}[1]{\left\langle #1 \right\rangle}

\def\be{\begin{equation}}
\def\ee{\end{equation}}
\def\beq{\begin{eqnarray}}
\def\eeq{\end{eqnarray}}

\newcommand{\C}[1]{\mathcal{#1}}
\begin{document}

\begin{center}
\vspace{48pt}
{ \Large \bf Instantons and Extreme Value Statistics of Random Matrices}

\vspace{40pt}
{\sl Max R. Atkin}$\,^{a}$,
and {\sl Stefan Zohren}$\,^{b}$  
\vspace{24pt}

{\small

$^a$~Fakult\"{a}t f\"{u}r Physik, Universit\"{a}t Bielefeld,\\
Postfach 100131, D-33501 Bielefeld, Germany

\vspace{10pt}

$^b$~Department of Physics, Pontifica Universidade Cat\'olica do Rio de Janeiro,\\
Rua Marqu\^es de S\~ao Vicente 225, 22451-900 G\'avea, Rio de Janeiro, Brazil.\\
}

\end{center}

%\addtolength{\baselineskip}{0.20\baselineskip}
%\vspace{2cm}

\vspace{36pt}

\begin{center}
{\bf Abstract}
\end{center}
We discuss the distribution of the largest eigenvalue of a random $N\times N$ Hermitian matrix. Utilising results from the quantum gravity and string theory literature it is seen that the orthogonal polynomials approach, first introduced by Majumdar and Nadal, can be extended to calculate both the left and right tail large deviations of the maximum eigenvalue. This framework does not only provide computational advantages when considering the left and right tail large deviations for general potentials, as is done explicitly for the first multi-critical potential, but it also offers an interesting interpretation of the results. In particular, it is seen that the left tail large deviations follow from a standard perturbative large $N$ expansion of the free energy, while the right tail large deviations are related to the non-perturbative expansion and thus to instanton corrections. Considering the standard interpretation of instantons as tunnelling of eigenvalues, we see that the right tail rate function can be identified with the instanton action which in turn can be given as a simple expression in terms of the spectral curve. From the string theory point of view these non-perturbative corrections correspond to branes and can be identified with FZZT branes.

\vspace{12pt}
\vfill

{\footnotesize
\noindent
$^a$~{email: matkin@physik.uni-bielefeld.de}\\
$^b$~{email: zohren@fis.puc-rio.br}\\
}
\newpage
\tableofcontents
\newpage

\section{Introduction}
Since their conception by Wigner in the 1950s, random matrices have found an ever growing number of applications in fields as diverse as statistical mechanics, quantum gravity, telecommunications and finance \cite{GernotsBook}. One reason for this vast range of applications is the phenomenon of universality, which refers to the fact that many properties of random matrices are independent of the measure used to define them. To be precise, let us consider a $N \times N$ random Hermitian matrix $M$ defined by the Gibbs measure,
\beq
\label{MMdef}
d\mu(M) = \frac{1}{\mathrm{Vol}[U(N)] Z_N}e^{-N \mathrm{Tr} V(M)} [dM],
\eeq
where $[dM]$ is the standard measure on Hermitian matrices, $V(M)$ is a finite degree polynomial known as the potential, $\mathrm{Vol}[U(N)]$ is the volume of the group $U(N)$ and $Z_N$ is the normalisation constant known as the partition function. As is well known, the limiting behaviour of the eigenvalues as $N \rightarrow \infty$ is characterised by them being confined to a finite number of finite length intervals on the real axis with a density given by the spectral density function $\rho(x)$. In the bulk of the eigenvalues the spectral density itself depends on the details of $V$ and therefore its macroscopic behaviour does not display any universality. However at the edge of the eigenvalue support, the spectral density generically has a square root behaviour independent of $V$ and thus displays a simple example of universality. 

A somewhat more famous example of universality arises when we consider $\mathbb{P}_N\left(\lambda_\mathrm{max}<z\right)$; the probability that the maximum eigenvalue is less than some given $z \in \mathbb{R}$. The probability distribution for small fluctations around the limiting largest eigenvalue is given in terms of the Tracy-Widom distribution, 
\beq
\lim_{N\rightarrow \infty}\mathbb{P}_N\left(c N^{2/3}(\lambda_\mathrm{max}-a)<z\right) =\exp\left(-\int_z^{+\infty}(y-z)q_0^2(y)dy\right),
\eeq
where $a$ is the edge of the eigenvalue support, $c$ is a potential dependent constant and $q_0$ is the Hastings McLeod solution to the Painlev\'{e} II equation. This distribution is obtained independently of $V$ apart from on a measure zero subset of potentials known as multi-critical potentials. The Tracy-Widom distribution itself has appeared in a huge number of areas \cite{TWreview} and therefore has attracted much attention over the years. One recent strain of investigation \cite{TC1,TC2,Adler2,HigherTW} has been the fate of the Tracy-Widom distribution when focusing on the above mentioned subset of multi-critical potentials. These non-generic potentials are characterised by the spectral density with a compact support of a single interval no longer having a square root behaviour at the support edge $a$, instead near the edge one has a behaviour of the form,
\beq
\rho(x) \propto (a-x)^{2k+1/2},
\eeq
where $k$ is an integer referred to as the order of multi-criticality. Such potentials are often referred to as multi-critical since if one considers the random matrix as a statistical mechanical system, they can be thought of as corresponding to phase transitions. To obtain such a multi-critical potential it suffices to make the replacement $V(M) \rightarrow \frac{1}{t} V(M)$ in \eqref{MMdef} and then choose $V(M)$ to have multiple minima. By varying $t$ one can then find a critical value $t_c$ at which a multi-critical behaviour is seen.\footnote{Another way of obtaining a different behaviour for the distribution of the largest eigenvalue is by considering a spectral density with non-compact support. This appears for example in Cauchy random matrices \cite{cauchy1,cauchy2}. In this case the instanton interpretation is slightly more subtle and we will not peruse it further here.} Another more rigorous way \cite{TC1, TC2} is to simply set the spectral density to the required form, i.e. $\rho(x) = c (x-b)^{1/2}(a-x)^{2k+1/2}$ and compute the potential using the saddle point equation,
\beq\label{PV}
\frac{1}{2}V'(x) = \mathrm{p.v.}\int^a_b dz \frac{\rho(z)}{x-z},
\eeq
where $\mathrm{p.v.}$ denotes the principle value. From this the form of an entire family of multi-critical potentials can be obtained \cite{TC2}.

For multi-critical potentials the Tracy-Widom distribution is modified to a $k$ dependent law first given in \cite{TC1} which were named higher order analogues of the Tracy-Widom distribution. These results were later rederived in a more compact form in \cite{HigherTW} using orthogonal polynomial techniques which generalised those of \cite{SN}.

\begin{figure}[t]
\centering 
%\parbox{7cm}{
\includegraphics[scale=0.7]{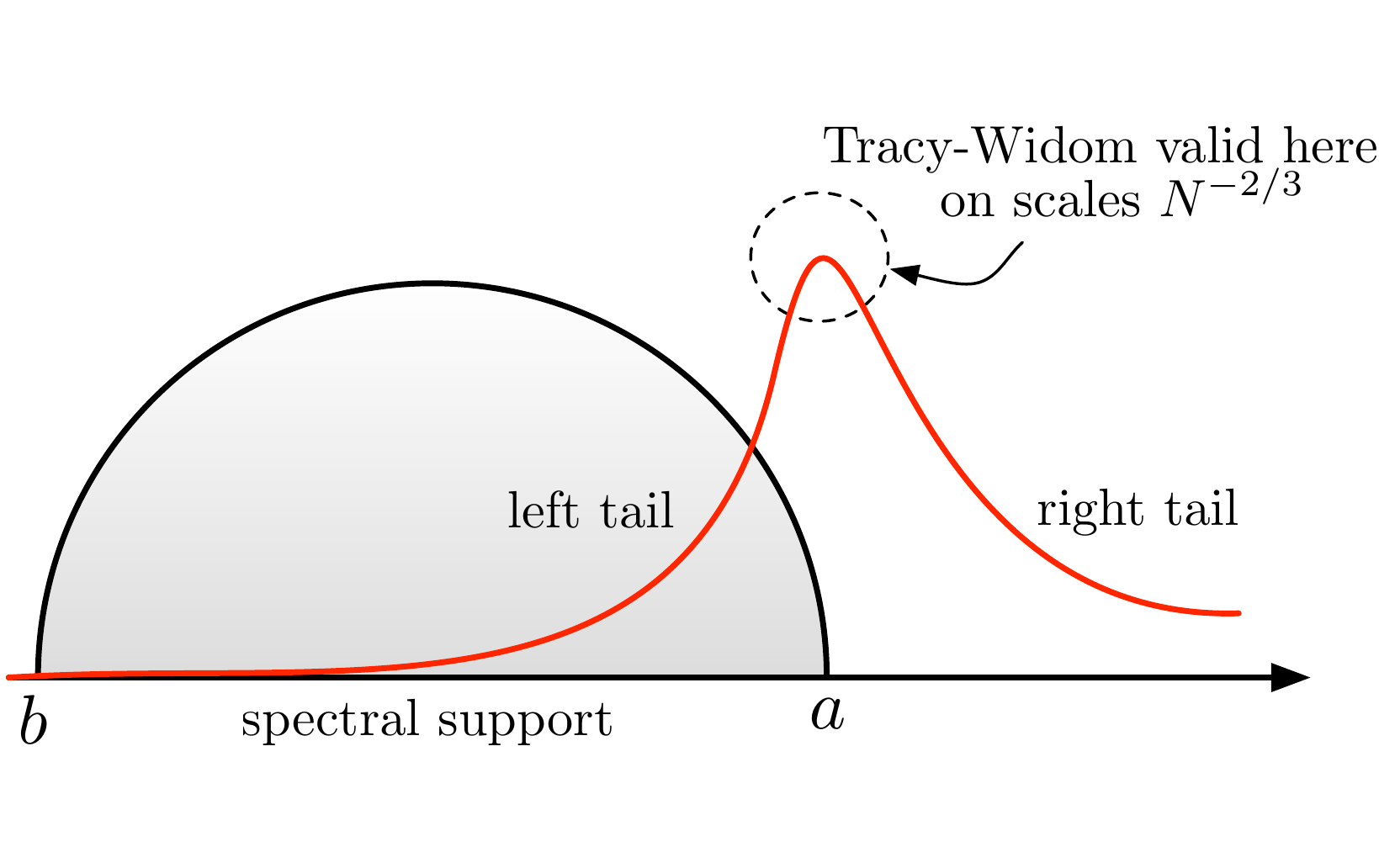}
\caption{A sketch of the probability distribution for the largest eigenvalue is shown in blue. Near the edge of the spectral support (the shaded region) the probability takes the form of the Tracy-Widom distribution. In this paper we are interested in the behaviour far out in the left and right tails.}
\label{fig1}
%}
\end{figure}

The Tracy-Widom distribution describes the ``typical'' fluctuations of the largest eigenvalue which were shown to occur on scales of $N^{-2/3}$ in the case of a generic potential. However, in recent years there has been a growing interest \cite{LargeDevSM1,LargeDevSM2,LargeDevSM3,LargeDevSM4,LargeDevTR1,LargeDevTR2} in the behaviour of ``atypical'' eigenvalues which appear very far from the limiting largest eigenvalue. This interest is motivated in part by applications to the landscape problem in string theory and to properties of random gaussian surfaces which have relevance to condensed matter \cite{LargeDevInString,LargeDevGlass,LargeDevGlass2}. A cartoon of the expected distribution is shown in Figure \ref{fig1} and we can see that the left and right tails that describe the large deviations are expected to be qualitatively different. This is because the deviations to the left must necessarily cause a rearrangement of the eigenvalues in the limiting bulk spectrum, whereas no such rearrangement is necessary for a single eigenvalue to appear far to the right. This has been borne out by detailed calculations using coulomb gas techniques \cite{LargeDevSM1,LargeDevSM2,LargeDevSM4}, topological recursion methods based on loop equations \cite{LargeDevTR1,LargeDevTR2} and an approach based on orthogonal polynomials \cite{SN}. These calculations have generally been for random matrices with Gaussian potential, with most interest focused on the dependence on the Dyson index $\beta$. In this paper we will only be dealing with Hermitian random matrices for which $\beta = 2$. Collecting the known results for Gaussian Hermitian matrices we have at leading order, \cite{LargeDevSM1,LargeDevSM2,LargeDevSM4,LargeDevTR1,LargeDevTR2}
\bea
\frac{d}{dz} \mathbb{P}_N\left(\lambda_\mathrm{max}<z\right) = \left\{
     \begin{array}{lr}
       \exp(-2 N^2 \psi_-(z) + \ldots), &z < a \\
       \exp(-2 N \psi_+(z)  + \ldots), &z > a
     \end{array}
   \right. \label{scalingleftright}
\eea
where the functions $\psi_\pm$, known as the left and right rate functions, are given by,
\bea
\psi_{-}(z)&=&\frac{z^2}{3}-\frac{z^4}{108}-\sqrt{z^2+6} \frac{\left(z^3+15 z\right)}{108}- \frac{1}{2}\ln\left[\frac{\sqrt{z^2+6}+z}{\sqrt{2}}\right]+\frac{\ln 3}{2},\label{psi-lit}\\
\psi_{+}(z)&=&\frac{z \sqrt{z^2-2}}{2} +\ln\left[\frac{z-\sqrt{z^2-2}}{\sqrt{2}}\right].
\eea

An obvious question is; can expressions for the left and right tail be obtained for arbitrary potentials and hence for the multi-critical cases? It is the purpose of this paper to address this issue within the context of the orthogonal polynomial approach begun in \cite{SN}. We note here that the topological recursion approach of \cite{LargeDevTR1,LargeDevTR2} is in principle powerful enough to answer this question; however only explicit results were given for the Gaussian case in these studies. We will find in the course of this paper that the orthogonal polynomial approach can be related to the instanton effects studied in \cite{MarinoSaddle,MarinoPolynomials,MarinoMulticut,MarinoReview} and via this we make contact with some results of \cite{LargeDevTR2}. Another perspective on the work here is that there are three well developed approaches to computing in random matrix theory; saddle point methods, loop equations and orthogonal polynomials. In the context of large deviations both the saddle point method and the loop equation method have been fully developed, however the orthogonal polynomial method has only been considered in \cite{SN}. It is an aim of this paper to fully develop this approach.

The plan of this paper is as follows: In Section \ref{Sec2} we review the orthogonal polynomial approach to gap probabilities which are given by the free energy in the presence of a hard wall. We describe the string equations in absence and presence of this hard wall. In Section \ref{Sec3} we describe how to use this approach to obtain the large deviations of the maximum eigenvalue. It is seen that the large deviations in the left tail can be obtained from the perturbative expansion of the free energy, while for the large deviations in the right tail non-perturbative effects and thus instantons effects are relevant. We apply this machinery both to the case of the Gaussian matrix model as well as to a case of a multi-critical potential. Having made the link between large deviations in the right tail and instantons effects we proceed in Section \ref{Sec5} by giving an interpretation in terms of eigenvalue tunnelling. Calculating the eigenvalue tunnelling using a saddle point method, enables one to obtain an expression for the right tail in terms of the spectral curve for an arbitrary potential. Furthermore, this provides interesting links to a string theoretic picture. We summarise our results and conclude in Section \ref{Sec6}. Appendix \ref{app1} provides some useful results regarding properties of the family of multi-critical potentials considered in Section \ref{Sec3}.

%%%%%%%%%%%%%%%%%%%%%%%%%%%%%%%%%%%%%%%%%%%%%%%%%%%%%%%%%%%%%%%%%%%%%%%%%%%%%%%%%
\section{The Orthogonal Polynomial Approach to Gap Probabilities}\label{Sec2}
%%%%%%%%%%%%%%%%%%%%%%%%%%%%%%%%%%%%%%%%%%%%%%%%%%%%%%%%%%%%%%%%%%%%%%%%%%%%%%%%%

\subsection{Orthogonal Polynomials for Matrix Models in Presence of a Hard Wall}

First we recall from the standard matrix model literature (see e.g.\ \cite{GernotsBook,DiFrancesco:1993nw,Marino:2004eq} for reviews) that we may integrate out the angular degrees of freedom in the matrix $M$ to obtain an expression for the partition function related to the Gibbs measure \eqref{MMdef} only in terms of the eigenvalues,
\beq
\label{Zinf}
Z_N\equiv Z_N(\infty; t)  = \frac{1}{N!}\int^{\infty}_{-\infty} \prod^{N}_{i=1} \frac{d\lambda_i}{2\pi} \Delta(\lambda)^2  e^{-\frac{N}{t} V(\lambda_i)}.
\eeq
where $\Delta(\lambda)=\prod_{i<j}(\lambda_i-\lambda_j)$ is the standard Vandermonde determinant. The orthogonal polynomial approach proposed in \cite{SN} begins by noting that we may write the gap probability as,
\beq
\BB{P}_N(z ;t) \equiv \BB{P}_N(\lambda_{\mathrm{max}} < z ;t) = \frac{Z_N(z; t) }{Z_N(\infty; t )},
\eeq
where $Z_N(z; t)$ is the cut-off partition function;
\beq
\label{Zz}
Z_N(z; t)  = \frac{1}{N!}\int^{z}_{-\infty} \prod^{N}_{i=1} \frac{d\lambda_i}{2\pi}  \Delta(\lambda)^2  e^{-\frac{N}{t} V(\lambda_i)}.
\eeq
We then proceed by introducing a set of polynomials $\{\pi_n(x): n \in \BB{N}\}$ such that $\pi_n(x)$ is of order $n$ and they are orthonormal with respect to the inner product defined by,
\beq
\braket{\pi_n}{\pi_m} \equiv \int^z_{-\infty} \frac{dx}{2\pi} e^{-\frac{N}{t} V(x)} \pi_n(x)\pi_m(x),
\eeq
i.e. $\braket{\pi_n}{\pi_m} = \delta_{nm}$. Note that the coefficients in the expression for each $\pi_n$ depend on $z$, $t$ and the potential. It is then a standard result of random matrix theory that the Vandermonde determinant may be written in terms of these polynomials, thereby allowing us to evaluate the partition function. In particular, defining $h_n(z)$ by $\pi_n(x) = \frac{1}{\sqrt{h_n(z)}} x^n  + O(x^{n-1})$, we have,
\beq
\label{Znorm}
Z_N(z; t) = \prod^{N-1}_{i=0} h_i(z) = h_0(z)^N \prod^{N-1}_{i=1} r_i(z)^{N-i}
\eeq
where we have defined the ratios
\beq
\label{rndef}
r_n(z) = \frac{h_n(z)}{h_{n-1}(z)} \quad \mathrm{for} \quad  n\geq 1,
\eeq
note that we will often suppress the arguments of $r_n$ and $h_n$. In order to compute the $r_n$ one derives a set of recursion relations. To do this we first introduce one final piece of notation. Consider the operator $x$ acting on the space of polynomials which merely multiplies a polynomial by $x$. We define its matrix representation by, $Q_{nm} = \braket{\pi_n }{x \pi_m}$. Note that from the definition of the inner product the operator $x$ is self-adjoint and therefore we have that $Q$ is symmetric. Furthermore, note that $\braket{\pi_{n+k} }{x \pi_n} = 0$ for $k > 1$, allowing us to conclude that $Q$ can be written as,
\beq
Q_{nm} = \sqrt{q_{n+1}}\delta_{n+1,m} + s_n\delta_{nm} + \sqrt{q_{n}} \delta_{n-1,m}
\eeq
where $q_n$ and $s_n$ are some yet to be determined functions $z$, $t$, and the potential. In fact by considering the quantity $\braket{\pi_n}{x \pi_{n-1}}$ and $\braket{\pi_n}{x \pi_{n}}$ one can show $q_n = r_n$ and $s_n = -\frac{g}{N} \partial_{t_1} \log h_n$ respectively.

In applications of random matrix theory to string theory, in which recursion relations for $r_n$ were first obtained, these recursion relations go by the name of string equations. In \cite{SN} recursion relations for $r_n$ were obtained in the case of a Gaussian potential. Unfortunately they had the unpleasant property of containing derivatives of $z$ which makes them difficult to use for potentials of arbitrary order. One of the advances made in \cite{HigherTW} was to obtain a set of purely algebraic recursion relations for any potential, which is what we will refer to as the string equations. Before reviewing the form of these string equations we will first consider the case of no hard-wall, i.e. when $z = \infty$.

%%%%%%%%%%%%%%%%%%%%%%%%%%%%%%%%%%%%%%%%%%%%%%%%%%%%%%%%%%%%%%%%%%%%%%%%%%%%%%%%%
\subsection{String Equations in Absence of a Hard Wall}
%%%%%%%%%%%%%%%%%%%%%%%%%%%%%%%%%%%%%%%%%%%%%%%%%%%%%%%%%%%%%%%%%%%%%%%%%%%%%%%%%

For the case of no hard wall the partition function can be determined from the string equations,
\beq
\label{streqnyinf}
[Q,P] = 1.
\eeq
The operator $P$ is defined as,
\beq
\label{Pfinal}
P = -\frac{N}{2t} (V'(Q)_+ - V'(Q)_-).
\eeq
where $\pm$ refer to the upper and lower triangular parts of the matrices.

This string equation can be used in a number of ways. Recursion relations for $r_n$ can be obtained from consideration of the diagonals, i.e. we have the recursion relations $[Q,P]_{nn} = 1$ and $[Q,P]_{n,n-1} = 0$. Alternatively, if we are only interested in the double scaling limit then we can take such a limit directly on the operators $Q$ and $P$. 
It is useful to note that the two non-trivial equations obtained from \eqref{streqnyinf} have the form,
\beq
[Q,P]_{n,n} -1 = (\sqrt{r_{n+1}}V'(Q)_{n+1,n}-\frac{t(n+1)}{N})-(\sqrt{r_{n}}V'(Q)_{n,n-1}-\frac{tn}{N}) = 0,
\eeq
and
\beq
[Q,P]_{n,n-1} =  V'(Q)_{n,n} - V'(Q)_{n-1,n-1},
\eeq
which are in turn implied by the integrated form of the string equations,
\beq
\sqrt{r_{n}}V'(Q)_{n,n-1} = \frac{tn}{N} \qquad V'(Q)_{nn} = 0.
\eeq

\subsection{String Equations in Presence of a Hard Wall}
In the presence of a hard wall the string equation becomes \cite{HigherTW},
\beq
\label{stringeqn}
[Q-z,H]_{nm}=(Q-z)_{nm},
\eeq
where $H$ is defined as,
\beq
H_{nm}&\equiv& (A(Q-z))_{nm} - \frac{N}{2t} (V'(Q)(Q-z))_{nm}
+\frac12\delta_{nm}
\label{Hdef}
\eeq
where $A$ is the matrix representation of the operator $\partial_x$.
It is also important to note that $H$ can also be expressed as \cite{HigherTW}
\beq
H=-\frac{N}{2t}\Big( (V'(Q)(Q-z))_+- (V'(Q)(Q-z))_-\Big),
\label{Hfinal}
\eeq
where the subscripts $\pm$ refer to the strictly upper and lower triangular parts of a matrix and subscript $d$ refers to the diagonal entries. %We now make two observations which will be important for later calculations. 
Note that one of the string equations in \eqref{stringeqn} can be written in an integrated form as $H_{nn} = 0$ which yields
\beq
\label{IntStr1}
 (V'(Q)(Q-z))_{nn} = \frac{(2n+1)t}{N}.
\eeq

%%%%%%%%%%%%%%%%%%%%%%%%%%%%%%%%%%%%%%%%%%%%%%%%%%%%%%%%%%%%%%%%%%%%%%%%%%%%%%%%%
\section{Large Deviations from String Equations and Instantons}\label{Sec3}
%%%%%%%%%%%%%%%%%%%%%%%%%%%%%%%%%%%%%%%%%%%%%%%%%%%%%%%%%%%%%%%%%%%%%%%%%%%%%%%%%

\subsection{Perturbative and Non-perturbative Expansion of the String Equations}

In \cite{SN} the right tail rate function together with its first correction term were obtained for a Gaussian potential using the orthogonal polynomial approach. The method introduced there, which was found to only work for the right tail, consisted of solving the recursion relations for $r_n$ by way of an ansatz\footnote{See equation (48) of \cite{SN}.} of the form,
\beq
r_n(z,t) = \frac{tn}{N} + c_n e^{-N z^2/t} + \ldots \qquad \mathrm{and} \qquad s_n(z,t) = d_n e^{-N z^2/t} + \ldots,
\eeq
which after plugging into the recursion relations resulted in recursion relations for $c_n$ and $d_n$ that can then be solved. It turns out that the large deviations to the right of the eigenvalue support is then governed by the term containing $c_n$. We now would like to draw attention to two aspects of the above ansatz. Firstly, both expressions consist of a power series in $N$, which in the case of $s_n$ is trivial, together with a term exponentially suppressed in $N$, plus potentially more terms which are even more strongly suppressed. Secondly the power series portion of the above ansatzes is identical, at finite $z$, to the exact solution when $z=\infty$. With these observations in mind we make the following claim; {\emph{one should think of the behaviour of the right large deviations as being due to a $z$ dependent non-perturbative correction to the $z = \infty$ solution}}. Here non-perturbative means that the $z$ dependence will not appear in the power series portion of the ansatz but only in the terms with exponential suppression as $N \rightarrow \infty$. This above claim is in a sense very much in the spirit of the early work by Gross and Matytsin \cite{Gross} which served as a motivation for \cite{SN}. Furthermore, it agrees with the observations made in \cite{weak1,weak2} that the right tail corresponds to the weak coupling regime.

In quantum field theory such non-perturbative corrections to perturbative expansions arise due to extra saddle points in the path integral which in turn correspond to solutions of the Wick rotated field equations. Such solutions go by the name instantons. In fact such instantons can also appear in random matrix theories, such as the class under consideration here and their computation at finite $N$ has been worked out in some detail \cite{MarinoReview}. In fact three distinct methods have been presented; one based on a saddle point analysis of the one-cut solution \cite{MarinoSaddle}, one on the multicut solutions \cite{MarinoMulticut} and one based on orthogonal polynomials \cite{MarinoPolynomials}. We will begin by applying the last of these to our current problem. 

% \cite{MarinoSaddle,MarinoPolynomials,MarinoMulticut,MarinoReview}

First let us review the procedure for computing the perturbative and non-perturbative corrections to $r_n$ when $r_n$ satisfies some set of recursion relations. In both cases one starts with the following steps:

\begin{itemize}
\item[(a)] In the one-cut case we can make the replacement $r_n \rightarrow r(tn/N)$ where $r(tn/N)$ is a smooth function of $tn/N$.
\item[(b)] The string equations now can be written as difference equations, since $r_{n+k} \rightarrow r(\zeta + t/N)$ where $\zeta = tn/N$. We will sometimes use the notation $r(\zeta ; z)$ rather than just $r(\zeta)$ when we want to emphasise the $z$ dependence.
\end{itemize}

The perturbative expansion, which as we will see is necessary to obtain the large deviation of the left tail, can be computed as follows:

\begin{itemize}
\item[(c)] The difference equation obtained in the previous step (b), which is called the pre-string equation in \cite{MarinoPolynomials}, can be solved perturbatively via substitution of the ansatz,
\beq
r(\zeta) = \sum^\infty_{k=0} g_s^{k} R^{(0)}_{k}(\zeta) \qquad s(\zeta) = \sum^\infty_{k=0} g_s^{k} S^{(0)}_{k}(\zeta),
\eeq
where we have introduced the useful parameter $g_s = t/N$.
\end{itemize}

In order to compute the non-perturbative corrections, which will be relevant to obtain the large deviations of the right tail, we modify this procedure,

\begin{itemize}
\item[(c')]  One starts by noting that the difference equation obtained in (b) admits a trans-series solution, which is a formal solution of the form,
\beq
\label{CSeries}
r(\zeta) = \sum^\infty_{l=0} C^{l} R^{(l)}(\zeta,g_s) \qquad s(\zeta) = \sum^\infty_{l=0} C^{l} S^{(l)}(\zeta,g_s),
\eeq
with,
\beq
R^{(0)}(\zeta) = \sum^\infty_{k=0} g_s^{k} R^{(0)}_{k}(\zeta)
\eeq
and 
\beq
\label{Rlan}
R^{(l)}(\zeta) = e^{-l A(\zeta)/g_s} \sum^\infty_{k=0} g_s^{k} R^{(l)}_{k}(\zeta)
\eeq
with similar expressions for $S$. The quantity $A(\zeta)$ is known as the instanton action and $R^{(l)}(\zeta)$ as the $l$'th instanton correction.
\item[(d')]  We make use of the above ansatz by substituting \eqref{CSeries} into the string equation. The coefficient of each power of $C$ gives a new string equation for the quantities $R^{(l)}(\zeta)$ and $S^{(l)}(\zeta)$. Into these new string equations we can substitute the ansatz \eqref{Rlan}.
\end{itemize}

Since the perturbative expansion is rather straightforward, we focus in the rest of this section on several details of the prescription to obtain the non-perturbative expansion needed to calculate the large deviations of the right tail:

When applying the above procedure we could work at the level of the recursion relations obtained from \eqref{stringeqn}, instead we will attempt to formulate the calculation in a way that allows an easier generalisation to any potential. We begin with \eqref{masteqn} and first substitute in the ansatz \eqref{CSeries}. This will lead to both $Q$ and $P$ being expanded in powers of $C$;
\beq
Q = \sum^\infty_{k=0} C^k Q^{(k)} \qquad P = \sum^\infty_{k=0} C^k P^{(k)}.
\eeq
Defining $\Delta = Q_+ - Q_-$ we will also have,
\beq
\Delta = \sum^\infty_{k=0} C^k \Delta^{(k)} \qquad V'(Q)_d = \sum^\infty_{k=0} C^k V'(Q)_d^{(k)}.
\eeq
We note that the operator $H$ defined in \eqref{Hdef} can be rewritten as \cite{HigherTW},
\beq
H&=& (Q-z)P -(QP)_{\rm d} -(Q_+-Q_-)\frac{N}{2t}V'(Q)_{d}\nn\\
&=& \frac12\{Q,P\}-zP-\frac{N}{4t}\{Q_+-Q_-,V'(Q)_{d}\},
\label{Hexp}
\eeq
and that the string equation \eqref{stringeqn} can be written using the above results as,
\beq
\label{masteqn}
\{Q-z,[Q,P]-1\} - [Q,\{Q_+-Q_-, \frac{N}{2t} V'(Q)_d \}] = 0.
\eeq
At zero'th order in C we obtain for \eqref{masteqn},
\beq
\{Q^{(0)}-z,[Q,P]^{(0)}-1\} - [Q^{(0)},\{\Delta^{(0)}, \frac{N}{2t} V'(Q)_d^{(0)} \}] = 0,
\eeq
which can be satisfied if we choose $Q^{(0)}$ and $P^{(0)}$ to have the exact same form as the case when we have $z = \infty$. What this amounts to is that when $z$ is finite there exists a solution to the string equations with the same perturbative solution as when $z = \infty$. In fact looking at the form of \eqref{masteqn} we see that it may be satisfied at any order of $C$ by choosing $Q^{(l)}$ and $P^{(l)}$ to have the same form as the $z=\infty$ case. However, due to the extra structure of \eqref{masteqn} there exist new $z$ dependent solutions. This $z$ dependent instanton contribution can be computed by looking at the next order in $C$, for which we have,
\beq
\label{masteqn2}
\{Q^{(0)},[Q^{(0)},P^{(1)}] + [Q^{(1)},P^{(0)}]\} - [Q^{(0)},\{\Delta^{(0)}, \frac{N}{2t} V'(Q)_d^{(1)} \}] = 0,
\eeq
where we have used the fact that $V'(Q)_d^{(0)} = 0$.

Now, the $Q$ and $\Delta$ operators are independent of the potential and can be written as,
\beq
Q^{(0)} = \sqrt{R^{(0)}(\zeta + t/N)} e^{\frac{t}{N} \partial} + S^{(0)}(\zeta) + \sqrt{R^{(0)}(\zeta)} e^{-\frac{t}{N} \partial},
\eeq
\beq
Q^{(1)} = \frac{\sqrt{R^{(1)}(\zeta + t/N)}}{2\sqrt{R^{(0)}(\zeta + t/N)}} e^{\frac{t}{N} \partial} + S^{(1)}(\zeta) + \frac{\sqrt{R^{(1)}(\zeta)}}{2\sqrt{R^{(0)}(\zeta)}} e^{-\frac{t}{N} \partial}
\eeq
and
\beq
\Delta^{(0)} = \sqrt{R^{(0)}(\zeta + t/N)} e^{\frac{t}{N} \partial} - \sqrt{R^{(0)}(\zeta)} e^{-\frac{t}{N} \partial}
\eeq
where $\partial \equiv \frac{\partial}{\partial \zeta}$. Of course the $P$ operator is dependent on the potential and therefore will have to be computed on a case-by-case basis and we will demonstrate this shortly by considering the case of a Gaussian potential. However, before proceeding to the example we will first consider how to compute the behaviour of the tails, once an expression for $r_n$ is obtained. We will do this by following the calculation in \cite{MarinoPolynomials}, making only minor alterations.

Using \eqref{Znorm} we may write,
\beq
\log \BB{P}_N(z ;t) = \log\left(\frac{h_0(z)}{h_0(\infty)}\right) - N \sum^{N-1}_{i=1} \left(1- \frac{i}{N}\right)\log \left(\frac{r_i(z)}{r_i(\infty)}\right)  
\eeq
which can be rearranged using the Euler-Maclaurin formula, into the form,
\beq
\label{Fdiffeqn}
F(z;t+g_s) + F(z;t-g_s) - 2F(z;t) = \log\left(\frac{r(t;z)}{r(t;\infty)}\right) 
\eeq
where $F(z;t) \equiv \log \BB{P}_N(z ;t)$. If we now substitute an ansatz for $F$, of the form,
\beq
\label{FCSeries}
F(z;t) = \sum^\infty_{l=0} C^{l} F^{(l)}(z;t,g_s)
\eeq
where
\beq
F^{(0)}(z;t,g_s) = \sum^\infty_{k=0} g_s^{k} F^{(0)}_{k}(z;t) \qquad \mathrm{and} \qquad F^{(l)}(z;t,g_s) = e^{-l A(t)/g_s} \sum^\infty_{k=0} g_s^{k} F^{(l)}_{k}(z;t)
\eeq
into \eqref{Fdiffeqn} we obtain,
\beq
\label{F0eqn}
F^{(0)}(z;t+g_s) + F^{(0)}(z;t-g_s) - 2F^{(0)}(z;t) = \log\left(\frac{R^{(0)}(t;z)}{r(t;\infty)}\right) 
\eeq
and
\beq
\label{Fleqn}
F^{(l)}(z;t+g_s) + F^{(l)}(z;t-g_s) - 2F^{(l)}(z;t) =\left[\frac{R^{(l)}(t;z)}{R^{(0)}(t;z)}\right]_c
\eeq
where the subscript $c$ denotes the connected piece,
\beq
\left[\frac{R^{(l)}(t;z)}{R^{(0)}(t;z)}\right]_c \equiv \sum^\infty_{s =1} \frac{(-1)^{s-1}}{s} \sum_{l_1 + \ldots l_s = l} \frac{R^{(l_1)}(t;z)}{R^{(0)}(t;z)} \ldots \frac{R^{(l_s)}(t;z)}{R^{(0)}(t;z)}.
\eeq
The ansatz for $R^{(l)}$ for $l>0$ leads to an expansion of the form,
\beq
\left[\frac{R^{(l)}(t;z)}{R^{(0)}(t;z)}\right]_c = e^{-l A(t)/g_s} \sum^\infty_{k=0} g_s^{k} c_{l,k+1}(z;t),
\eeq
and it can be shown \cite{MarinoPolynomials} that \eqref{Fleqn} can be solved order by order with the result,
\beq
\label{Fl1eqn}
F^{(l)}_{1}(z;t) = \frac{1}{4} c_{l,1} \mathrm{cosech}^2 \left( l A'(t)/2 \right).
\eeq
The solution for higher order $F^{(l)}_{k}(z;t)$ can be found in \cite{MarinoPolynomials}. We now have all the technical tools at hand to tackle the computation of the large deviations in the left and right tail in a number of examples.

%%%%%%%%%%%%%%%%%%%%%%%%%%%%%%%%%%%%%%%%%%%%%%%%%%%%%%%%%%%%%%%%%%%%%%%%%%%%%%%%%
\subsection{Large Deviations in the Case of Gaussian Potential}
%%%%%%%%%%%%%%%%%%%%%%%%%%%%%%%%%%%%%%%%%%%%%%%%%%%%%%%%%%%%%%%%%%%%%%%%%%%%%%%%%

%%%%%%%%%%%%%%%%%%%%%%%%%%%%%%%%%%%%%%%%%%%%%%%%%%%%%%%%%%%%%%%%%%%%%%%%%%%%%%%%%
\subsubsection{Large Deviations in the Left Tail}
%%%%%%%%%%%%%%%%%%%%%%%%%%%%%%%%%%%%%%%%%%%%%%%%%%%%%%%%%%%%%%%%%%%%%%%%%%%%%%%%%

We first focus on the large deviation of the left tail in the case of a Gaussian matrix model. As already mentioned above, we will see that to obtain the large deviation of the left tail we only need to calculate the \emph{perturbative expansion} of $Z_N(z; t, \{g_k\})$. This is done by using the string equations in presence of a hard wall. Let us note that for the Gaussian potential to first order the potential dependent operator is given by,
\beq
P^{(0)} = -\frac{N}{2t}(\sqrt{R^{(0)}(\zeta + t/N)} e^{\frac{t}{N} \partial} - \sqrt{R^{(0)}(\zeta)} e^{-\frac{t}{N} \partial}),
\eeq
The first integrated string equation, \eqref{IntStr1} implies,
\beq
2 R^{(0)}_0(\zeta, z) - z S^{(0)}_0(\zeta, z) + S^{(0)}_0(\zeta, z)^2 =2 \zeta \label{Guassianleft-tmp}
\eeq
and secondly \eqref{stringeqn} with $m = n$ yields
\beq
\frac{\partial}{\partial\zeta} \left( R^{(0)}_0(\zeta, z)( 2 S^{(0)}_0(\zeta, z) -z )    \right) =S^{(0)}_0(\zeta, z)-z
\eeq
where the differential comes from the difference of two expressions at consecutive values of $N$.
This may be integrated by using \eqref{Guassianleft-tmp} resulting in
\beq
S^{(0)}_0(\zeta, z)^2 (z^2  - 4 \zeta - 4 z  S^{(0)}_0(\zeta, z) + 3   S^{(0)}_0(\zeta, z)^2) = 0.
\eeq
We see that the solution has three distinct branches
\beq
S^{(0)}_0(\zeta, z)=0,\quad S^{(0)}_0(\zeta, z) =\frac{2 z  \pm \sqrt{ z^2  +12 \zeta }}{3  }.
\eeq
If for the moment we consider the solution as a function of $z$ we know that for $z\to\infty$ we must have the solution $S^{(0)}_0(\zeta, z)=0$, which corresponds to $R^{(0)}_0(\zeta, z) =R^{(0)}_0(\zeta, \infty)= \zeta$. However, since this solution is constant it must eventually change to a different branch for smaller values of $z$. This happens continuously at $z=2\sqrt{\zeta}$, where it intersects the solution with the minus in front of the square root. We thus have
\beq
R^{(0)}_0(\zeta ,z) =\begin{cases}
     \zeta & \zeta\leq \zeta_c(z), \\
    \frac{6 \zeta +z \left(z  +\sqrt{z^2 +12 \zeta}\right)}{18}  &\zeta\geq\zeta_c(z) ,
\end{cases} \label{eq-Rcase}
\eeq
where $\zeta_c(z)= z^2/4$. Hence, we have for the left tail
\bea
g_s^2 F^{(0)}(z;t) &=&  \int_0^t d\zeta (t-\zeta) \log \left( \frac{R^{(0)}_0(\zeta,z)}{R^{(0)}_0(\zeta;\infty)} \right)\nn\\
&=& \int_{\zeta_c(z)}^t d\zeta (t-\zeta) \log \left( \frac{R^{(0)}_0(\zeta,z)}{R^{(0)}_0(\zeta;\infty)} \right), \label{eq:Pleft-i}
\eea
since in the interval $\zeta\in[0,\zeta_c(z)]$ the integrand is zero by \eqref{eq-Rcase}.
There thus is a critical value for $z$ for which $\zeta_c(z)=t$ and which is given by $z_c=2\sqrt{t}$. Hence, we have
\beq
g_s^2F^{(0)}(z;t) =  
\begin{cases}
       \int_{\frac{z^2}{4}}^t d\zeta (t-\zeta) \log \left( \frac{6 \zeta +z \left(z +\sqrt{  z^2  +12 \zeta }\right)}{18 \zeta}  \right) & z\leq 2\sqrt{t}, \\
     0  &  z\geq 2\sqrt{t}.
\end{cases}
\eeq
This yields to leading order in the large $N$ expansion,
\beq
g_s^2F^{(0)}(z;t) =  \begin{cases}
      -\frac{tz^2}{3}  +\frac{z^4}{216} +\frac{z}{216} (z^2+30t) \sqrt{z^2+12 t} + t^2 \log\left(\frac{z+\sqrt{z^2+12 t}}{6\sqrt{t}}\right)  & z\leq 2\sqrt{t}, \\
     0  &  z\geq 2\sqrt{t}.
     \end{cases} \label{eq:Pleft-f}
\eeq
where we recall that $g_s=t/N$. This expression reproduces the correct result from the literature \cite{LargeDevSM1,LargeDevSM2}, i.e.\ equation \eqref{psi-lit}. One observes that the logarithm of the probability function $\log \BB{P}_N(z ;t) $ is zero for $z\geq 2\sqrt{t}$, i.e.\ to the right of the cut, when calculated at order $N^2$ in the perturbative expansion. This reflects the different scaling of the large deviations in the left and right tail: it is easier to pull a single eigenvalue out of the bulk than it is to push it into the bulk as one has to rearrange other eigenvalues as well. This different behaviour is manifest in the scaling, i.e.\ \eqref{scalingleftright}. From a statistical physics point of view one sees that the above behaviour corresponds to a phase transition at $z= 2\sqrt{t}$ in the limit $N\to\infty$.

Alternatively to the derivation above, leading from \eqref{eq:Pleft-i} to \eqref{eq:Pleft-f}, we could have also used the differential version of this equation 
\bea
g_s^2 \partial^2_t F^{(0)}(z;t) & =&  \log \left( 6 t +z \left(z +\sqrt{  z^2  +12 t }\right) \right) -\log (18 t)\nn\\
&=&  2 \log \left(z +\sqrt{  z^2  +12 t }\right) -\log (18 t) \label{eq:P-left-diff}
\eea
which follows directly from \eqref{F0eqn}. To obtain the probability function from \eqref{eq:P-left-diff} we have to integrate it twice. This gives rise to two integration constants (which are function of $z$) in the final expression, $f_0(z) +f_1(z) t$. These can be fixed as follows: From the fact that $R^{(0)}_0(t,z=2\sqrt{t})=R^{(0)}_0(t,z=\infty)$, we see that the logarithm of the probability function $F^{(0)}(z;t)=0$  for $z\geq2\sqrt{t}$. The same should be true for the probability density, i.e. $\partial_z \BB{P}^{(0)}_N(z ;t)=0$  for $z\geq2\sqrt{t}$. However, $[\partial_z \BB{P}^{(0)}_N(z ;t)]_{z=2\sqrt{t}}=0$ does also imply $[\partial_t \BB{P}^{(0)}_N(z ;t)]_{z=2\sqrt{t}}=0$. This gives us two conditions $[F^{(0)}(z;t)]_{z=2\sqrt{t}}=0$ and $[\partial_t F^{(0)}(z;t)]_{z=2\sqrt{t}}=0$. Both conditions determine $f_0(z)= 7 z^4/216$ and $f_1(z)=-z^2/3$ which fixes the solution which agrees with the previous derivation \eqref{eq:Pleft-f}.

%%%%%%%%%%%%%%%%%%%%%%%%%%%%%%%%%%%%%%%%%%%%%%%%%%%%%%%%%%%%%%%%%%%%%%%%%%%%%%%%%
\subsubsection{Large Deviations in the Right Tail}
%%%%%%%%%%%%%%%%%%%%%%%%%%%%%%%%%%%%%%%%%%%%%%%%%%%%%%%%%%%%%%%%%%%%%%%%%%%%%%%%%

As is seen in the previous subsection the large deviations of the right tail are zero when calculating $\log \BB{P}_N(z ;t) $ at order $N^2$ using the perturbative expansion. In this case the \emph{non-perturbative expansion} will be relevant. Hence we have to include the next term of order $N$. To do so we note that the potential dependent operators to next order are,
\beq
P^{(1)} = -\frac{N}{2t}(\frac{\sqrt{R^{(1)}(\zeta + t/N)}}{2\sqrt{R^{(0)}(\zeta + t/N)}} e^{\frac{t}{N} \partial} - \frac{\sqrt{R^{(1)}(\zeta)}}{2\sqrt{R^{(0)}(\zeta)}} e^{-\frac{t}{N} \partial})
\eeq
and
\beq
V'(Q)_d^{(1)} = S^{(1)}(\zeta).
\eeq
We also recall from the previous subsection that the perturbative solution is $R^{(0)}(\zeta) = \zeta$, $S^{(0)}(\zeta) = 0$ for the right tail. If we now substitute these operators into \eqref{masteqn2}, expand in powers of $g_s$ and equate the coefficients of each differential operator to zero, we obtain the following,
\beq
2 e^{A'(\zeta)} \left(-z+\sqrt{\zeta}+\sqrt{\zeta} \cosh\left(A'(\zeta )\right)\right)R^{(1)}_0(\zeta ,z)+\left(1+e^{A'(\zeta)}\right) \left(2 \zeta -z   \sqrt{\zeta}\right)S^{(1)}_0(\zeta ,z)=0
\eeq
and
\beq
zS^{(1)}_0(\zeta ,z)- \left(1+e^{-A'(\zeta)}\right) R^{(1)}_0(\zeta ,z) = 0.
\eeq
This leads to,
\beq
4 \zeta \cosh(A'(\zeta)/2)^2 = z^2.
\eeq
Furthermore, by expanding up to order $g_s^2$ we can obtain, with the help of Mathematica, the following relations;
\beq
S^{(1)}_1(\zeta ,z) = \frac{e^{-A'(\zeta)} \left(2 R^{(1)}_1(\zeta ,z)+2 e^{A'(\zeta)} R^{(1)}_1(\zeta ,z)-R^{(1)}_1(\zeta ,z) A''(\zeta)+2 \partial_\zeta R^{(1)}_1(\zeta ,z)\right)}{2 z}
\eeq
and
\beq
S^{(1)}_2(\zeta ,z) = \frac{\left(\left(1+e^{-A'(\zeta)}\right) R^{(1)}_1(\zeta ,z)+2 \left(1-e^{-A'(\zeta)}\right) \zeta \left(\left(1+e^{A'(\zeta)}\right) R^{(1)}_2(\zeta ,z)+\partial_\zeta R^{(1)}_1(\zeta ,z)\right)\right)}{2 \left(-1+e^{A'(\zeta)}\right) z \zeta }.
\eeq
This leads finally to,
\beq
R^{(1)}_1(\zeta ,z) =  \kappa_1 \frac{\sqrt{\zeta }}{\sqrt{z^2-4 \zeta }}
\eeq
where $\kappa_1$ is a constant of integration. In order to reproduce the right large gap asymptotics of the Tracy-Widom distribution we must choose $\kappa_1 = 0$ and in this case we find further that,
\beq
R^{(1)}_2(\zeta ,z) =  \kappa_2 \frac{\sqrt{\zeta }}{\sqrt{z^2-4 \zeta }},
\eeq
where again, $\kappa_2$ is a constant of integration. If we now substitute these results into \eqref{F0eqn} we find,
\beq
F^{(0)}(z;t+g_s) + F^{(0)}(z;t-g_s) - 2F^{(0)}(z;t) = 0
\eeq
which upon expanding in powers of $g_s$ gives $F^{(0)}(z;t) = 0$. Additionally, using \eqref{Fl1eqn} we have,
\beq
F^{(1)}_1(z;t) = 0
\eeq
and
\beq
F^{(1)}_2(z;t) = \kappa_2 \frac{\sqrt{t}}{(z^2-4t)^\frac{3}{2}}.
\eeq
Putting these into \eqref{Fleqn} gives,
\beq
\log \BB{P}_N(z ;t) = \frac{\kappa_2}{N} \left(\frac{t}{z^2-4t}\right)^\frac{3}{2} e^{-\frac{N A(t)}{t}} + \ldots  
\eeq
with
\beq
A(t) = \frac{1}{2} z^2 \sqrt{1-\frac{4 t }{z^2}}-t  \mathrm{ArcCosh}\left[-1+\frac{z^2}{2 t }\right],
\eeq
thus reproducing the known result obtained in \cite{SN} up to an overall constant. Unfortunately, within the instanton approach taken here the constant $\kappa_2$ can not be fixed; see \cite{MarinoPolynomials}. This is because the trans-series approach only produces a family of formal solutions parameterised by such unknown constants. Among the members of such a family of trans-series solutions there exist the 'true' solutions to the difference equation however the particular values of the constants corresponding to such solutions can only be found via a non-perturbative definition of the theory, which in this case is the original matrix integral. However, it is possible to make some general remarks regarding such constants. In principle the above analysis does not forbid such constants being $z$ dependent however it is known that any asymptotic expansion of a solution to a difference equation holds in extended region of the complex plane known as a Stoke sector. This means that within such a region the constants are independent of $z$. The boundaries between such regions are known as Stoke lines and here we may see a discontinuous jump in the values of the constants as move from one region to the next. This effect is known as the Stokes phenomenon and we will comment on this effect again later in the paper. Finally let us explain why in \cite{SN} they were able to to obtain the value for the unknown constant. The analysis in \cite{SN} was significantly more complicated than that performed here as the authors in fact solve the recursion relation exactly rather than just for large $N$. They could then choose the unknown constant to reproduce the explicitly calculated form of $r_1$.

\subsection{Large Deviations in the Case of a Multi-critical Potential}
A set of multi-critical potentials which yield convergent matrix integrals are \cite{TC2},
\beq
\label{MCV}
V_k(x) = \frac{2(2k+2)!}{(-1/2)_{2k+2}} \sum^{2k+1}_{l=0}\frac{(-1)^l(l+1/2)_{2k+1-l}}{((2k+1-l)!(l+1))}x^{l+1},
\eeq
where $(a)_l$ is the Pochhammer symbol. These potentials produce a random matrix whose spectral density has support on $[0,1]$ and has the form
\beq
\rho(x) \propto x^{1/2} (1-x)^{2k+1/2}.
\eeq
which can be checked explicitly by e.g.\ using \eqref{PV}. It is worthwhile noticing that the family of multi-critical potentials results in convergent matrix integrals. Furthermore, as is shown in Appendix \ref{app1} for any $k\in\mathbb{N}$ the potential $V_k(x)$ only possess a single minimum for $x$ on the real axis. In the following we will actually consider a generalisation of the above multi-critical potentials by setting our potential to $V(x) = \frac{1}{t} V_k(x)$. In this section we will use the tools from the previous section to compute the rate functions in the case $k = 1$ for which we have
\beq
\label{multiV}
V_1(x) =16\left( -x +3x^2 -\frac{8}{3} x^3 +\frac{4}{5}x^4 \right).
\eeq

\subsubsection{Large Deviations in the Left Tail}

We proceed as in the case of the Gaussian potential. Firstly, the integrated string equation, \eqref{IntStr1} implies,
\bea
%\frac{16}{5} \left(96 R^2+(2 S (15+4 S (2 S-5)) -5) (S-z)+4 R (12 S (4 S-2 z-5)+5 (3+4 z))\right)=2\zeta, \label{stringk1-1}
\frac{16}{5} \left(60 R+96 R^2-5 S-240 R S+30 S^2+192 R S^2-40 S^3+16 S^4+5 z+\right.\nn\\
\left.+ 80 R z-30 S z-96 R S z+40 S^2 z-16 S^3 z\right)=2\zeta, \label{stringk1-1}
\eea
where we used the shorthand notation $R\equiv R^{(0)}_0(\zeta, z)$ and $S\equiv S^{(0)}_0(\zeta, z)$.
Secondly \eqref{stringeqn} with $m = n$ yields,
\bea
%\frac{\partial}{\partial_\zeta} \left(  \frac{16}{5} R \left(24 R (8 S-2 z-5)-5 (1+6 z)+4 S (2 S (8 S-6 z-15)+5 (3+4 z) ) \right)  \right)=S-z.\label{stringk1-2}
% \frac{\partial}{\partial_\zeta} \left( 
%-16 R-384 R^2+192 R S+\frac{3072}{5}R^2 S-384 R S^2+\frac{1024}{5}R S^3+\right. \nn\\
%\left. -96 R z-\frac{768}{5}R^2 z+256 R S z-\frac{768}{5} R S^2 z
 %\right)=S-z.\label{stringk1-2}
  \frac{16}{5} \frac{\partial}{\partial\zeta} \left( 
-5 R-120 R^2+60 R S+192 R^2 S-120 R S^2+\right. \nn\\
\left. +64 R S^3-30 R z-48 R^2 z+80 R S z-48 R S^2 z
 \right)=S-z.\label{stringk1-2}
\eea
This equation can in principle be integrated by using \eqref{stringk1-1} to eliminate $R$ in \eqref{stringk1-2} analogous to the Gaussian case. However, the detailed algebraic manipulation is cumbersome and not very instructive. We therefore pay our attention only to the universal behaviour which can be deduced from a simple argument. Recall, the expression of the rate function of the left tail large deviations in the Gaussian case, i.e.\ \eqref{eq:Pleft-f}. The detailed form of this expression changes in general when considering a different potential, however, the behaviour close to the end point $a$ of the spectral support is universal for potentials which have a spectral density with square root behaviour. In particular, an expansion of \eqref{eq:Pleft-f} around $z=a\equiv 2\sqrt{t}$ yields,
\beq
g_s^2F^{(0)}(a-\omega;t) \sim \omega^3.
\eeq
Instead of expanding the final result, we could have also obtained this result from \eqref{eq:Pleft-i}, or alternatively  \eqref{eq:P-left-diff}, by expanding $R^{(0)}_0(\zeta,z)/R^{(0)}_0(\zeta,\infty)$ around $\zeta=\zeta_c(z)=z^2/4$,
\beq
\frac{R^{(0)}_0(\zeta,z)}{R^{(0)}_0(\zeta;\infty)}  =1-\frac{2}{z^2}(\zeta-\zeta_c(z))+...
\eeq
yielding the above result. In particular, using $\zeta_c(a-\omega)=t- \zeta_c'(a)\omega+...$ we observe the scaling
\beq
R^{(0)}_0(\zeta,a-\omega)=R^{(0)}_0(\zeta;\infty)(1- \frac{2}{a^2}\left(\zeta - t +\zeta_c'(a)\omega\right) +... ).
\eeq
This suggests that in the case of the multi-critical potential we can make a scaling ansatz very much in the spirit of the scaling ansatz made in the case of the double scaling limit used to obtain the higher-order Tracy-Widom distributions \cite{HigherTW}. Notice that in \cite{HigherTW} the double scaling limit is taken directly on the string equations, i.e.\ \eqref{stringeqn} which are the same equations used to obtain \eqref{stringk1-1} and \eqref{stringk1-2}. Hence, it is not surprising that we can deduce the general behaviour from this analysis. In particular, we will now argue that in the case of the higher multi-critical potentials one has 
\beq
\frac{R^{(0)}_0(\zeta,z)}{R^{(0)}_0(\zeta;\infty)}  =1-g(\zeta)\, (z_c(\zeta)-z)^{4k+1}+...=1-g(\zeta_c(z)) \, (\zeta-\zeta_c(z))^{4k+1}+...
\eeq
which implies
\beq
g_s^2F^{(0)}(a-\omega;t) \sim\int_{t-\zeta_c'(a) \omega}^t d\zeta (t-\zeta)(\zeta-t+ \zeta_c'(a) \omega)^{4k+1}\sim \omega^{4k+3}.
\eeq
In \cite{HigherTW} it is shown that to obtain a double scaling limit in which $F^{(0)}(a-\omega;t)$ remains finite as $\omega\to 0$, the string equations \eqref{stringeqn} imply that one has to scale (see also \cite{Bowick:1991ky})
\beq
N^{-1} \sim g_s\sim \omega^{\frac{4k+3}{2}}
\eeq
leading to the above result and is in accordance with previous works \cite{TC1}. Furthermore, the same analysis also implies that the right tail rate function has to scale near the edge of the eigenvalue support as the square root of the corresponding expression for the left tail (see for example \eqref{scalingleftright}) and thus $A(t,a+\omega)\sim \omega^{(4k+3)/2}$, as we will see in later sections.

\subsubsection{Large Deviations in the Right Tail}

The operators $P^{(0)}$, $P^{(1)}$ and $V'(Q)^{(1)}_d$ can be computed from the formula \eqref{Pfinal} in the standard way; we do not reproduce the exact expression here as they are too large to be illuminating. Using the fact that the perturbative portion of $Q$ and $P$ are unchanged by the presence of the hard wall we obtain the following relations,
\beq
\label{r0sol}
-\zeta +96 R+\frac{768}{5} R{}^2-256 R S+\frac{768}{5} R S{}^2 =0,
\eeq
\beq
-16-256 R+96 S+\frac{1536}{5} R S-128 S{}^2+\frac{256}{5} S{}^3 =0,
\eeq
where $R = R^{(0)}_0(\zeta ,z)$ and $S = S^{(0)}_0(\zeta ,z)$. The second of these equations can be rearranged to obtain an expression for $R^{(0)}_0(\zeta ,z)$ as a rational function of $S^{(0)}_0(\zeta ,z)$. Eliminating $R^{(0)}_0(\zeta ,z)$ from the above equations also gives an expression for $\zeta$ in terms of $S^{(0)}_0(\zeta ,z)$,
\beq
\zeta = \frac{-135+1300 S-4620 S^2+8192 S^3-7808 S^4+3840 S^5-768 S^6}{(-5+6 S)^2}.
\eeq
If we now substitute the standard large $N$ ansatz into \eqref{masteqn2}, expand in powers of $g_s$ and equate the coefficients of each differential operator to zero, we obtain the following,
\bea
S^{(1)}_1(\zeta ,z)&=& \frac{1}{K} 2e^{-A'(\zeta)} \left(1+e^{A'(\zeta)}\right) (-65+330 S-520 S^2+256 S^3- 100 z+ \nn\\
&& +240 S z-144 S^2 z+ 5-30S+40 S^2-16 S^3 \cosh\left(A'(\zeta)\right)) R^{(1)}_1(\zeta ,z),
\eea
where
\bea
K&=& 25-50 S-80 S^2+240 S^3-128 S^4-130 y+460 S z-560 S^2 z +\nn \\
&&+224 S^3 z+\left(-5+30 S-40 S^2+16 S^3\right) (-5+8 S-2 z) \cosh\left(A'(\zeta)\right),
\eea
where we have used \eqref{r0sol} to eliminate $R^{(0)}_0(\zeta ,z)$. We also obtain the equation,
\beq
\label{ASeqn}
&&-5+30 S-80 S^2+64 S^3+80 S z-96 S^2 z-40 z^2+48 S z^2+\left(-5+30 S-40 S^2+16 S^3\right)\times\nn \\
&& \cosh\left[\frac{(-5+6 S)^3 \frac{dA}{dS}}{16 (-1+2 S)^2 \left(305-1180 S+1740 S^2-1152 S^3+288 S^4\right)}\right] = 0,
\eeq
which can be integrated to give,
\bea
A(\zeta;z)&=&\frac{2}{15} (-2 \sqrt{(a-z) (b-z)}(45 a^3+45 b^3+3 a^2 (-40+9 b-6 z)-6 b^2 (20+3 z)+\nn \\
&&b \left(90+80 z-24 z^2\right)-4 z \left(45-40z+12 z^2\right)+ \nn \\
&&a \left(90+27 b^2+80 z-24 z^2-4 b (20+3 z)\right))- \nn \\
&&3 (a-b)^2 \left(15 a^2+2 a (-20+9 b)+5 \left(6-8 b+3 b^2\right)\right) \text{ArcCosh}\left[-\frac{a+b-2
z}{a-b}\right]),
\eea
where we have introduced $a = S^{(0)}_0(\zeta ,z) + 2 \sqrt{R^{(0)}_0(\zeta ,z)}$ and $b = S^{(0)}_0(\zeta ,z) - 2 \sqrt{R^{(0)}_0(\zeta ,z)}$, which correspond to the end points of the spectral support when $\zeta = t$.

In fact \eqref{ASeqn} is not the only manner in which the conditions arising from the expansion of \eqref{masteqn2} may be satisfied. The equation obtained from the expansion of \eqref{masteqn2} that gives $A$ consists of a number of factors, one of which must be zero. Be requiring the only $z$ dependent factor to be zero we obtain \eqref{ASeqn} however there are other $z$ independent factors which yield distinct instanton actions, which we will denote $\tilde{A}$. This gives the trans-series solution,
\beq
r(\zeta;z) = R^{(0)}(\zeta;\infty) + Ce^{-A(\zeta;z)/g_s}(R^{(1)}_1(\zeta ,z) + \ldots) + \tilde{C}e^{-\tilde{A}(\zeta)/g_s}(\ldots) + \ldots
\eeq
where we have introduced a separate set of unknown constants $\tilde{C}$ for the $z$ independent instanton. If we now substitute the above into \eqref{F0eqn} and \eqref{Fleqn} we obtain,
\beq
\label{TCrighttail}
\log \BB{P}_N(z ;t) = Ce^{-A(\zeta;z)/g_s}(\ldots) + \tilde{C}e^{-\tilde{A}(\zeta)/g_s}(\ldots)  + \ldots
\eeq
where we have permitted ourselves to redefine the arbitrary constants when necessary. Again we stress that the trans-series approach does not fix these constants. Here it appears that this may cause problems since we now have two possible instanton corrections that are competing and we would like to know their relative strengths. However, we note that the problem of competing corrections does not affect the (leading order) of the probability density function, for which,
\beq
\partial_z \BB{P}_N(z ;t) \sim e^{-A(\zeta;z)/g_s}.
\eeq
We will see in the next section that using some information from the matrix integral we can easily make more detailed statements concerning the unknown constants appearing in the trans-series. Before moving on to that discussion let us briefly consider the behaviour of the rate function near the edge of the eigenvalue support. It is known that the right tail rate function in the case of the Tracy-Widom distribution has the following behaviour \cite{SN, TC1},
\beq
A(t; a + \omega) \sim \omega^{3/2}.
\eeq
If we expand the rate function computed for the multi-critical potential about the edge of the eigenvalue support we obtain,
%\beq
%A(t; a + \omega) = \frac{64}{15} \sqrt{a-b} \left(15 (-1+a)^2+2 (-5+3 a) b+3 b^2\right) \omega ^{3/2}+O[\omega ]^{5/2}.
%\eeq
%Recalling that $a$ and $b$ are both functions of $t$ it is a simple exercise to check that the coefficient of the $\omega^{3/2}$ term vanishes when $t=1$, which is precisely when the potential reduces to the multi-critical potential $V_1$. In this case the behaviour of the rate function at the edge of the support is,
\beq
A(t; a + \omega) \sim \omega^{7/2}.
\eeq
As we will see below the general behaviour for the $k$-th multi-critical potential is $A(t; a + \omega) \sim \omega^{(4k+3)/2}$.

\section{Eigenvalue Tunnelling and the Spectral Curve}\label{Sec5}
%** First introduce saddle point methods <-- mention that we can interpret the rate function as the instanton describing tunneling up to the hard wall. 
%** Then apply to TC multicritical potential to show that the hard wall gives the only instanton correction. This requires us to discuss the steepest desent contours
%** Then discuss arbitrary potential and the qualitative behaviour

\subsection{Eigenvalue Tunnelling and the Qualitative Behaviour of $\BB{P}_N(z ;t)$}
In the preceding sections we have used the trans-series approach to extract the rate function from the orthogonal polynomial formulation. This approach has the advantage that one can obtain the tail behaviour very quickly in comparison to the analysis in \cite{SN} but with the disadvantage that the coefficients determining the strength of the corrections are left unknown. As was mentioned previously, there exist alternative approaches to computing the instanton action which involve a steepest descent analysis of the matrix integral, such an approach has also been used to compute the rate function \cite{LargeDevSM1}. Since the purpose of this article is to push the techniques of \cite{SN} further and apply them to the multi-critical potentials, our focus will not be on redoing the above calculations in the saddle point approach, in order to compute the unknown constants. Instead we will look into the saddle point calculations in \cite{MarinoSaddle} in order to gain more insight into the qualitative behaviour of $\BB{P}_N(z ;t)$.

The instantons of the matrix models considered in \cite{MarinoSaddle} are generally associated to the process of ``eigenvalue tunnelling''. In order to discuss this phenomenon further, let us follow \cite{MarinoSaddle} and make use of the fact that the integrand in \eqref{Zinf} is holomorphic and therefore we can deform the contour away from the real axis. Hence we can write the partition function without hard wall in the following form \cite{MarinoSaddle},
\beq
\label{Zgamma}
Z_N(\infty; t)  = \frac{1}{N!}\int_\gamma \prod^{N}_{i=1} \frac{d\lambda_i}{2\pi} \Delta(\lambda)^2  e^{-\frac{N}{t} V(\lambda_i)}.
\eeq
where $\gamma$ is now a contour in the complex plane beginning at $-\infty$ and ending at $+\infty$. We can choose $\gamma$ such that it is composed of a sequence of contours $\gamma_k$, each of which begin and end at infinity and approach it in a direction such that $e^{-\frac{N}{t} V(\lambda)}$ decays exponentially. Furthermore one can choose the $\gamma_k$ to be steepest descent contours each passing through a unique saddle point $x_k$ of the integrand, i.e $V'(x_k) = 0$ (see \cite{Felder:2004uy}). However, note that $\gamma$ might not pass through all saddle points of the integrand. Assuming $\gamma$ passes through $d$ saddle point, we can write the partition function in the form,
\beq
Z_N(\infty; t) = \sum_{N_1 + \ldots + N_d = N} Z(N_1,\ldots,N_d)
\eeq
where 
\beq
Z(N_1,\ldots,N_d) = \frac{1}{\prod^d_{k=1} N_k!}\prod^d_{k=1}\left(\int_{\lambda^{(k)}_i \in \gamma_k}  \prod^{N_k}_{i=1} \frac{d\lambda^{(k)}_i}{2\pi}  e^{-\frac{N}{t} V(\lambda^{(k)}_i)}\right) \Delta(\lambda)^2,
\eeq
where $\Delta(\lambda)$ is the standard Vandermonde determinant evaluated on all eigenvalues. The above integral corresponds to separating the eigenvalues in \eqref{Zgamma} into groups of size $N_k$, which are only integrated along the $\gamma_k$ contour. The functions $Z(N_1,\ldots,N_d)$ are sometimes known as ``backgrounds'', being denoted by $(N_1,\ldots,N_d)$, and one can gain some insight into their meaning by considering the limit $t\rightarrow 0$. In this limit the repulsion between eigenvalues is turned off and the eigenvalues $\lambda^{(k)}_i$ localise at $x_k$, the saddle point, on $\gamma_k$ \cite{MarinoSaddle}. In this limit we have,
\beq
Z(N_1,\ldots,N_d) \sim \prod^d_{k=1} \left( e^{-\frac{V(x_k)}{g_s}} \right)^{N_k}
\eeq
For non-zero $t$ the eigenvalues at the $k$-th saddle point will spread out to fill the neighourhood around $x_k$. The backgrounds in which all eigenvalues localise about a single saddle point are known as one-cut backgrounds and the one-cut assumption corresponds to assuming that the lowest energy configuration is such a background. If we change a one-cut background by moving a single eigenvalue to a new cut about the $k$-th saddle point, i.e.
\beq
(N,0,0,\ldots,0) \mapsto (N-1,\ldots,0,1,0,\ldots,0)
\eeq
then the partition functions are related approximately by,
\beq
\frac{Z(N-1,\ldots,0,1,0,\ldots,0)}{Z(N,0,0,\ldots,0)} \sim \exp\left[-\frac{N}{t}\left(V(x_k) - V(x_1)\right) \right]
\eeq
the above quantity can in fact be identified with the small $t$ limit of the instanton action \cite{MarinoSaddle} and therefore the instantons are often interpreted as describing the tunnelling of eigenvalues from one background to another.

Now let us consider how the above picture changes once a hard wall is introduced at $z$. The equation \eqref{Zgamma} is still valid but the contour $\gamma$ now begins at $-\infty$ and ends at $z$. Furthermore, by a trivial extension of the argument in \cite{Felder:2004uy} we can again decompose $\gamma$ into a number of contours $\gamma_k$ which possess the same properties as the no hard-wall case together with a contour joining $z$ to infinity and approaching infinity along a direction on which $e^{-\frac{N}{t} V(\lambda)}$ decays exponentially; see Figure \ref{fig4}. We can again choose this contour to be a steepest descent path i.e.\ a path on which the imaginary part of $V(\lambda_i)$ is constant. Because it is a steepest descent path we have that $V(\lambda)$ is monotonically decreasing along it and therefore attains its minimum value at $z$. The main contribution to the contour ending at $z$ hence comes from the region near $z$ and we have again the relation,
\beq
\frac{Z(N-1,0,\ldots,1)}{Z(N,0,0,\ldots,0)} \sim \exp\left[-\frac{N}{t}\left(V(z) - V(x_1)\right) \right]
\eeq
where $Z(N-1,0,\ldots,1)$ is the partition function when a single eigenvalue is present at the hard wall and the rest remain in the one-cut. \emph{In the language of instantons we would then identify the probability distribution for large deviations to the right, as being determined by the process of eigenvalue tunnelling up to the hard wall.}

\begin{figure}[t]
\centering 
%\parbox{7cm}{
\includegraphics[scale=0.7]{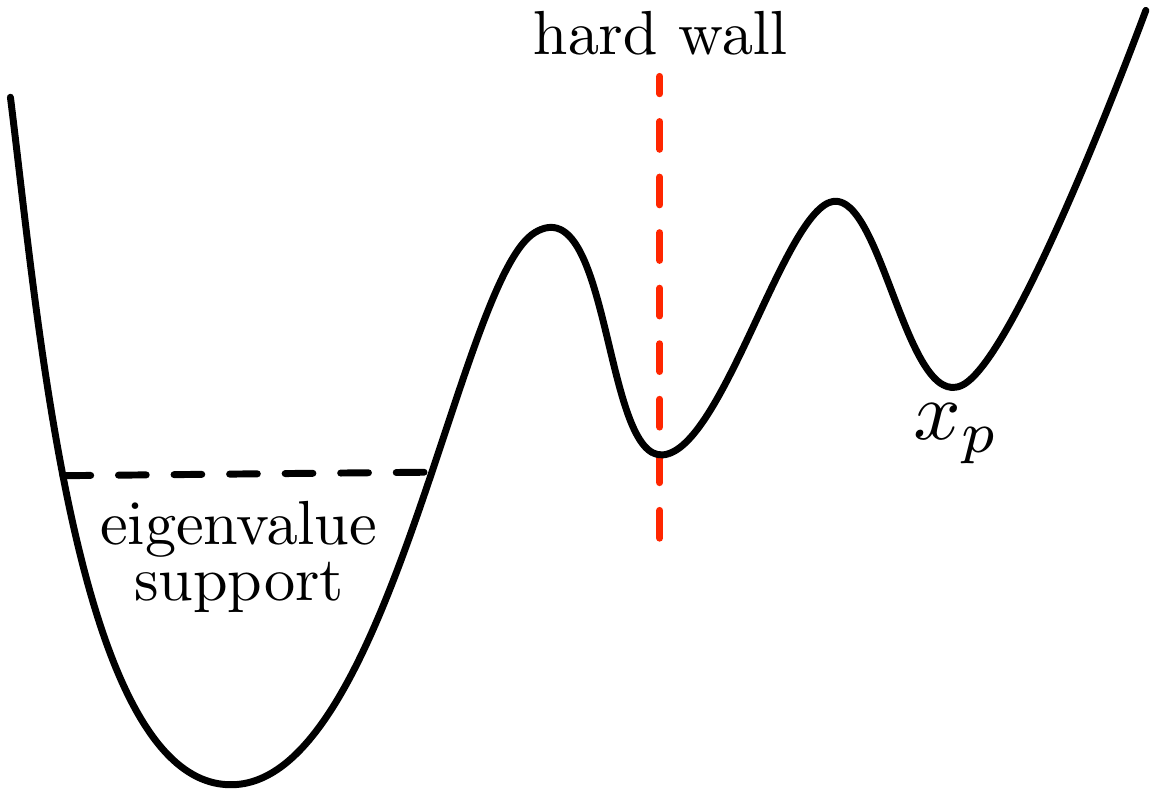}
\caption{An example of a general potential with multiple minima. The behaviour of the right tail in this case will be dominated by the process of eigenvalue tunnelling to the minimum at $x_p$ if the hard wall is placed somewhere in the intervening landscape, as shown.}
\label{fig2}
%}
\end{figure}

Finally, let us comment on the relation of the above picture to Stoke sectors. As we vary $z$, the contours composing $\gamma$ will change as $z$ moves to the right of further saddle points. When such a change in contours occurs, the trans-series solution will change discontinuously, which is precisely the Stokes phenomenon. We also note that the $z$ dependent instanton may not be the dominant contribution to the probability. Consider the situation in which we have a potential like that shown in Figure \ref{fig2}; one for which the eigenvalues reside in the deepest minimum on the left and there exists some landscape of minima to its right. Consider the case when the deepest minimum to the left of the hard wall position occurs at $x_p$, then the we may write the probability as,
\bea
\log \BB{P}_N(z ;t) &=& F(z;t) - F(\infty;t)\nn \\
  &\approx&  C_{h.w} e^{-\frac{V(z)}{g_s}}F^{(1)}_0(z; t|h.w) + \left(\sum_{k:x_k < z} C_k e^{-\frac{V(x_k)}{g_s}} F^{(1)}_0(\infty; t|k) + \ldots \right) \nn \\
&-&\left[C_p e^{-\frac{V(x_p)}{g_s}} F^{(1)}_0(\infty; t|p) +\ldots+ \left(\sum_{k:x_k < z} C_k e^{-\frac{V(x_k)}{g_s}} F^{(1)}_0(\infty; t|k) + \ldots \right) \right]
\eea
where in the second equality we have substituted in the trans-series and introduced the notation $F^{(l)}(\infty; t|k)$ to denote $F^{(l)}(\infty; t)$ associated with the $k$-th saddle point or the hard wall (h.w.). Note that the contribution from the saddle points to the left of the hard wall cancel and therefore it is only the saddle points to the right of the wall which effect the cumulative probability and this effect is dominated by the the deepest minimum of these, at $x_p$. We therefore have,
\bea
\label{sadP}
\log \BB{P}_N(z ;t) &=& F^{(1)}(z; t|h.w.) - F^{(1)}(\infty; t|p) + \ldots \nn \\
&=&  C_{h.w} e^{-\frac{V(z)}{g_s}}F^{(1)}_0(z; t|h.w.) - C_p e^{-\frac{V(x_p)}{g_s}}F^{(1)}_0(\infty; t|p) + \ldots
\eea
and we see there is a competition between the instanton for tunnelling to the hard wall and the instanton for tunnelling to the deepest minimum to the right of the wall. The picture is that the probability of finding an eigenvalue at a hard wall to the right of the support depends on $z$ until $V(z) > V(x_p)$ where $x_p$ is the position of the next deepest minima to the right of the support. When $V(z) > V(x_p)$ the large deviation probability is nearly independent of $z$ until $z$ reaches the saddle point at $x_p$. One can think of this as being due to the fact that the most likely place to find an eigenvalue far from the bulk is in the next deepest minimum. Hence the cumulative probability does not increase until we have included this minima in our integration.

\begin{figure}[t]
\centering 
%\parbox{7cm}{
\includegraphics[scale=1]{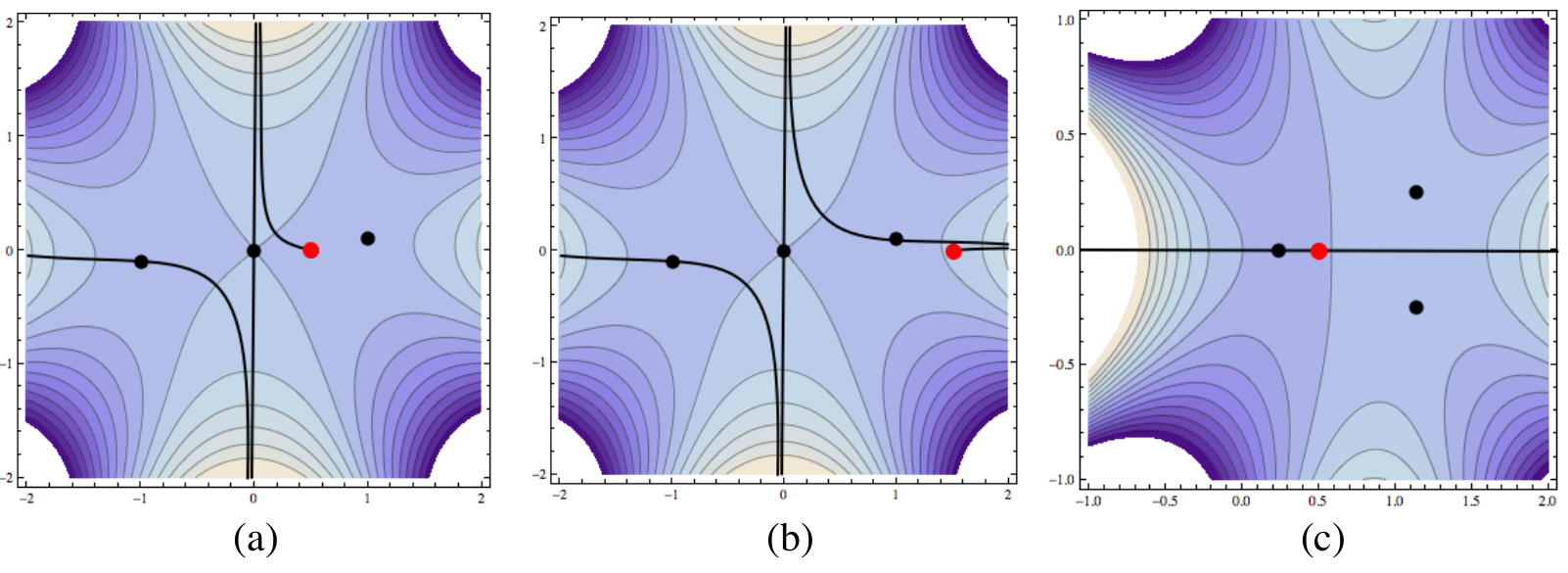}
\caption{A number of examples showing the contours $\gamma_k$ in the presence of a hard wall. These figures show a small section of the complex plane centered at the origin. We have plotted a contour plot of the real part of the potential function $V(z)$ together with a red dot at the location of the hard wall and black dots at the location of saddle points of $V$. The black lines follow the paths of steepest descent. Note that all steepest descent paths leave via the white asymptotic regions where the integral converges. The first two figures (a,b) are plotted using the potential defined such that $V'(x) = 3(x-1-i \epsilon)(x+1+i \epsilon)x$ with $\epsilon = 0.1$. This potential was chosen for the pedagogical reason that all steepest descent lines are distinct. We see that as the hard wall moves to the right from figure (a) to figure (b), the set of contours change to include the final saddle point. The final diagram (c) is for the multi-critical potential \eqref{multiV}. In this case we see that there is one saddle point at $z=0$ which contributes and the steepest descent contour leaving the saddle point to the right coincides with the path taken back from infinity to the hard wall.}
\label{fig4}
%}
\end{figure}

We now turn our attention to the multi-critical potentials \eqref{MCV}. These potential have only a single turning point on the real axis and therefore when no hard wall is present, the contour $\gamma$ in \eqref{Zgamma} coincides with the real axis, going from $-\infty$ to $\infty$. When a hard wall is present to the right of the cut, then $\gamma$ consists of the previous contour going from $-\infty$ to $+\infty$ along the real axis followed by a contour going from $+\infty$ back along the real axis to $z$; see Figure \eqref{fig4}. We note, in particular, that by the results of the Appendix \ref{app1} for no value of $z$ does $\gamma$ pass through the saddle points off the real axis. We can therefore conclude that there is no Stokes phenomenon for this particular potential and \eqref{TCrighttail} in fact holds with $\tilde{C} = 0$ and $C$ a $z$ independent constant. In the next section we will use the saddle-point approach of \cite{MarinoSaddle} to compute $C$.

%% Write down t=0 picture, in which eigenvalue accumulate about critical points
%% Sometimes the partition functions Z(N_1,\ldots,N_d) are referred to as backgrounds. Changing the background by moving a single eigenvalue changes the partition function by blah, which is precisely the small t form of the instanton action. Instanton are therfore interpreted as being due to the tunnelling of eigenvalues from one vacua to another.

%% How does this picture change once a hard wall is introduced? 
%% Repeat formulas with justification.
%% Need to show that still can choose gamma_i contours
%% Mention that we can interpret the right tail function as giving the contribution form eigenvalues tunneling up to the wall
%% Describe why we only get the z dependent instanton in the Claeys potential
%% Describe in general that we will get a competition of terms for a general potential

\subsection{The Rate Function and the Spectral Curve} 
We now consider how some of the results of \cite{MarinoSaddle} can be applied to compute the probability \eqref{sadP} including constants and relate this to the spectral curve. We begin by reviewing some standard facts about matrix models in order to fix notation and keep the discussion here self-contained. Firstly, we introduce the resolvent,
\beq
W(x) = \avg{\mathrm{Tr} \frac{1}{x-M}} = \sum^\infty_{g = 0}g_s^{2-2g-1} W_g(x),
\eeq
whose planar part, $\omega(x) = \frac{1}{t}W_0(x)$, satisfies the loop equations,
\beq
t \omega(x)^2 - V'(x) \omega(x) + p(x) = 0
\eeq
when no hard wall is present, where $p(x)$ is a polynomial in $x$ whose coefficient may be fixed using the one-cut assumption.  
It is customary to define $y(x) = V'(x) - 2t \omega(x) $ and think of the loop equations as defining a complex algebraic curve known as the spectral curve. The resolvent is related to the spectral density at large $N$ by,
\beq
\omega(x) = \int ds \frac{\rho(s)}{x-s}
\eeq
where the integral is over the support of $\rho$. Note that the resolvent and hence the spectral curve is unchanged when the hard wall is to the right of the spectral density support, since in this case the spectral density is identical to when no hard wall is present. In what follows it will also be useful to define what is known in the literature as the holomorphic effective potential (see e.g.\ \cite{MarinoReview}),
\beq
V_{\mathrm{h,eff}}(x) \equiv V(x) - 2t \int^x ds\,  \omega(s),
\eeq
note that $V_{\mathrm{h,eff}}'(x) = y(x)$. In \cite{MarinoSaddle} it is shown that,
\beq
F^{(1)}(\infty; t|k) = \frac{1}{2\pi} \frac{Z(N-1)}{Z(N)} \int_{\gamma_k} \exp \left(-\frac{1}{g_s} V_{\mathrm{h,eff}}(x) + \Phi(x) \right),
\eeq
where, $Z(N) \equiv Z(N,0,\ldots,0)$, i.e. the one-cut background containing $N$ eigenvalues and $\Phi(x)$ has a perturbative definition,
\beq
\Phi(x) = \sum_n g_s^n \Phi_n
\eeq
with $\Phi_1(X) = -\log\left[(x-a)(x-b)\right]$ where $b$ is the left of the eigenvalue support. The higher order terms $\Phi_n$ are computed in \cite{MarinoSaddle}. Finally it will be useful to write,
\beq
\frac{Z(N-1)}{Z(N)} = \exp \left(\sum_n g_s^n \C{G}_n(t) \right)
\eeq
where it is known that $\mathcal{G}_0(t) = V_{\mathrm{h,eff}}(a)$, as well as $\mathcal{G}_1(t) = \log\frac{a-b}{4}$ and the higher order terms are given in \cite{MarinoSaddle}.
We can write $F^{(1)}(\infty; t|p)$ as,
\beq
F^{(1)}(\infty; t|p) = \frac{1}{2 \pi} e^{\C{G}_1(t)} \int_{\gamma_k} dx  e^{-\frac{1}{g_s} \left(V_{\mathrm{h,eff}}(x) - V_{\mathrm{h,eff}}(a)\right) + \Phi_1(x)}(1 + O(g_s^2))
\eeq
With this expression in hand we can now proceed by the method of steepest descent to obtain,
\beq
\label{F1p}
F^{(1)}(\infty; t|p) = \sqrt{\frac{g_s}{2\pi V_{\mathrm{h,eff}}^{\prime\prime}(x_p)}}\exp\left(\Phi_1(x_p) + \C{G}_1(t)\right) e^{-\frac{A}{g_s}}
\eeq
where $A = V_{\mathrm{h,eff}}(x_p) - V_{\mathrm{h,eff}}(a) = \int^{x_p}_a y(x) dx$.
In the case of the hard wall the main contribution comes from the end point at $z$, so we obtain,
\beq
\label{F1hw}
F^{(1)}(\infty; t|h.w) = -\frac{g_s(a-b)}{8\pi y(z)(z-a)(z-b)} \exp\left(-\frac{1}{g_s}\int^z_a y(x) dx\right).
\eeq
Substituting \eqref{F1hw} and \eqref{F1p} into \eqref{sadP} yields an expression for the right deviation probability including the constants. Specialising to the case of the multi-critical potentials \eqref{MCV} we have for the right deviation probability ,
\beq
\log \BB{P}_N(z ;t)  = -\frac{g_s(a-b)}{8\pi y(z)(z-a)(z-b)} \exp\left(-\frac{1}{g_s}\int^z_a y(x) dx\right).
\eeq
We note also that we may reproduce the form of the Gaussian case obtained earlier, by setting $y(x) = \sqrt{x^2 - 4t}$ and noting $a = -b = 2 t^{1/2}$. This gives,
\beq
\log \BB{P}_N(z ;t) = -\frac{t^{3/2}}{2\pi N(z^2 - 4t)^{3/2}} \exp\left(-\frac{1}{g_s}\int^z_a y(x) dx\right).
\eeq
which is precisely the expression obtained in \cite{SN} after setting $t = \frac{1}{2 \alpha}$.

Furthermore, we see that the rate function for the right tail is given by the integral over the spectral curve
\beq
A(t,z)=\int^z_a y(x) dx.
\eeq
In the case of the multi-critical potentials \eqref{MCV} the spectral curve behaves as $y(x) \sim (x-a)^{2k+1/2}$ near the edge of the spectrum. Hence, one has
\beq
A(t,a+\omega)=\int^z_a y(x) dx \sim \omega^{2k+3/2}.
\eeq

\subsection{String Theoretic Comments}
It is well known that the matrix models considered in this paper have a double scaling limit which produces amplitudes of a continuum theory referred to as minimal string theory. Depending on the choice of multi-critical potential one obtains Liouville theory coupled to a minimal model CFT of $(p,q)$ type. The expansion in $g_s$ is then identified with the topological expansion of the string world sheet. Clearly in this picture the non-perturbative corrections we have been computing here correspond to non-perturbative stringy effects and indeed such effects have received a lot of attention in the literature; \cite{MarinoReview,ZZinstantons,ZZinstantons2MM,DavidMMInstantons} to highlight a few. We now wish to highlight some connections between the results here and some of the results of these previous studies.

\begin{figure}[t]
\centering 
%\parbox{7cm}{
\includegraphics[scale=0.7]{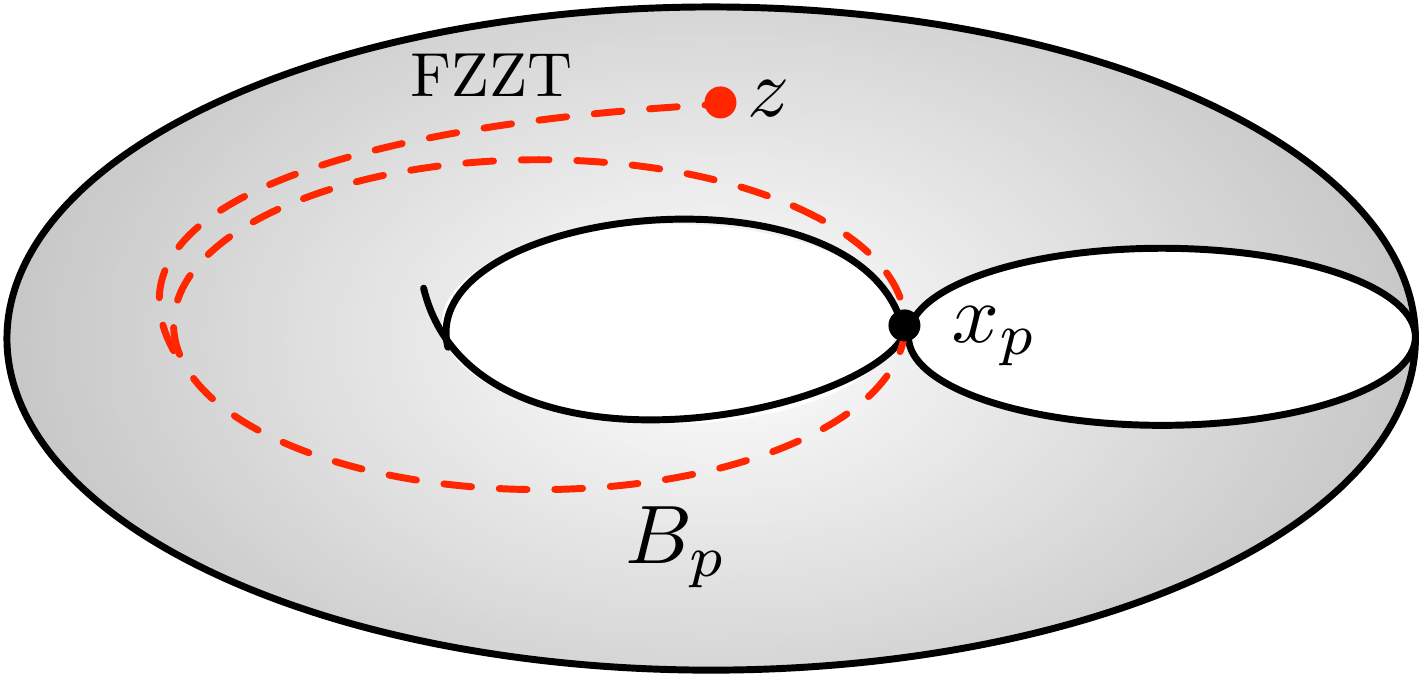}
\caption{The above figure shows the Riemann surface associated to the spectral curve $y(x)$. The support of the eigenvalue density corresponds to the one non-pinched cycle. The other minima of the potential, in which eigenvalues could in principal lie, correspond to pinched cycles. The ZZ brane disc function is computed by integrating the differential form $y(x)dx$ around the $B$-cycle passing through one of the pinched cycles. The FZZT brane corresponds to integrating the same form from an arbitrary point to a point $z$ depending on its boundary cosmological constant. The relation between the branes and integrations on the spectral curve was first uncovered in \cite{Seiberg:2003nm}}
\label{fig3}
%}
\end{figure}

Firstly, it is interesting to look at the string theoretic interpretation of the instanton corrections considered here. In particular the instanton action corresponding to eigenvalue tunnelling from the cut to another minima has the form,
\beq
A = \int^{x_p}_a y(x) dx.
\eeq
If we consider the spectral curve $y(x)$ and its associated Riemann surface we see it has the form shown in Figure \ref{fig3}, in which the absence of other eigenvalue cuts corresponds to the pinching of cycles. The instanton action can be rewritten in the form,
\beq
A = \frac{1}{2} \oint_{B_p} y(x) dx
\eeq
where the integration contour is now the B-cycle connecting the cut to the pinched cycle corresponding to the $p$th minima. In the double scaling limit such an integral is well known in the minimal string literature and has been identified with the disc function associated with a ZZ brane. Such disc functions corresponds to the world sheet partition function for a closed string to emerge from the bulk vacuum and annihilate with the brane. In this way the ZZ branes were identified as objects responsible for the non-perturbative corrections to the string world sheet expansion.

However, besides the ZZ branes there exists another class of branes in minimal string theory known as FZZT branes. Such branes have disc functions of the following form,
\beq
A = \int^{z} \tilde{y}(x) dx
\eeq
where $\tilde{y}$ is the double scaled spectral curve. We see that the FZZT brane disc function corresponds precisely to the double scaled limit of the instanton action for tunnelling up to the hard wall. As far as the authors are aware this is the first instance of FZZT branes playing the role of providing D-instanton like corrections to the worldsheet expansion. 

These observations mean that the double scaled limit of the theory with hard wall, which is precisely the limit needed to obtain the higher order analogue Tracy-Widom distributions, corresponds to a string theory with the same perturbative expansion as the usual theory but with extra non-perturbative corrections due to FZZT branes. This is particularly interesting in light of the appearance of Painlev\'{e} XXXIV in the context of type 0A string theory \cite{Klebanov:2003wg}. It would appear that type 0A string theory corresponds to precisely the bosonic theory with extra non-perturbative contributions from FZZT brane instantons.

\section{Conclusion}\label{Sec6}

In this paper we discuss  large deviations of the maximum eigenvalue of random $N\times N$ Hermitian matrices. 
In general such an analysis can be performed using saddle point methods, loop equations or orthogonal polynomials, where we mainly focus on the latter. The orthogonal polynomials approach to gap probabilities was first introduced by Majumdar and Nadal \cite{SN}, as described in the introduction, where it was used to calculate the right tail large deviations of the maximum eigenvalue in the case of a Gaussian potential. 

By restricting ourselves to Hermitian matrices we can make use of extensive results from the quantum gravity and string theory literature. It is seen how the approach of \cite{SN} can be embedded into a general algorithm for computing the left and right tail large deviations for a general potential from the modified string equations in presence of a hard wall, which were first introduced in \cite{HigherTW} and which are reviewed in Section \ref{Sec2}. This general framework is described in detail in Section \ref{Sec3} where it is also applied to explicitly compute the left and right tail large deviations in the case of a Gaussian ensemble, as well as for the first multi-critical case. Besides the computational advantages of this framework, one of the main results of this paper lies in its interpretation: It is seen that the left tail large deviations can be obtained from the perturbative large $N$ expansion of the free energy in presence of a hard wall, while the right tail large deviations are related to the non-perturbative expansion of the  free energy. This leads to the interesting observation that the right tail large deviations are related to instanton effects. More explicitly, it is shown that the right tail rate function is given by the instanton action. This is in accordance with an earlier observation that the right tail corresponds to the weak coupling regime \cite{weak1,weak2}. One disadvantage of the employed, so-called trans-series, approach is that the right tail large deviations probability function is only given up to $z$-independent terms. However, when calculating the probability density to leading order these terms drop out.

From the above analysis we have explicitly calculated the right tail rate function for the first multi-critical potential. For the general case of the $k$-th multi-critical potential we obtain the following behaviour close to the endpoint $a$ of the eigenvalue support: on the left one has $\frac{d}{dz}\log\BB{P}_N(z\!=\!a-\omega ;t)\sim \exp(-N^2 \omega^{4k+3})$, as $\omega\to0$ and on the right $\frac{d}{dz}\log\BB{P}_N(z\!=\!a+\omega ;t)\sim \exp(-N \omega^{(4k+3)/2})$, as $\omega\to0$. This behaviour was also obtained by an analysis of a Riemann-Hilbert problem \cite{TC1} (see also \cite{TC2}) and it reflects the fact that to obtain a nontrivial rescaled probability distribution in the central part, one has to scale $\omega\sim N^{-2/(4k+3)}$ which is the higher order analog of the scaling depicted in Figure \ref{fig1}. In particular, the above limiting behaviour is used to fix the boundary conditions of the central part, the double scaled probability distribution which corresponds to the higher analogs of the Tracy-Widom distribution and which is given in terms of solutions to the Painlev\'e II hierarchy \cite{TC1}.

Having established the relation between the right tail large deviations of the maximum eigenvalue and instantons of the random matrix ensemble we explore this relation further in Section \ref{Sec5}. Building on works by Mari\~no \cite{MarinoSaddle,MarinoPolynomials,MarinoMulticut,MarinoReview} in the string theoretic literature we discuss the interpretation of the above described instantons effects as an eigenvalue tunnelling from the bulk to the hard wall. Extending the saddle point analysis of \cite{MarinoSaddle} to include eigenvalue tunnelling from the bulk to the hard wall, we obtain a simple closed expression for the right tail large deviation probability in terms of the spectral curve including the overall constant. More explicitly, we see that the instanton action and thus the right tail rate function is simply given by the integral over the spectral curve from the end point of the cut to the position $z$ of the hard wall. This gives rise to a further interesting interpretation from a string theoretic point of view: It is well know that the instanton action corresponding to the tunnelling of an eigenvalue from the end point of the cut to another minimum outside of the bulk is related in the double scaling limit to the disc function associated with ZZ-branes. When introducing the hard wall the eigenvalue tunnels from the end point of the cut to hard wall located at $z$. We observe that in this case, the instanton action is related to the disc function associated with a FZZT brane instead. While the relation of instanton effects in the worldsheet theory and ZZ-branes is well studied (see for instance \cite{MarinoReview,ZZinstantons,ZZinstantons2MM,DavidMMInstantons}), as far as the authors are aware, this is the first appearance of a relation between FZZT branes and instanton effects in minimal string theory.

We hope to have convinced the reader that, as in the spirit of \cite{HigherTW}, the quantum gravity and string theory literature provides us with powerful techniques to study aspects of the distribution and large deviations of the maximum eigenvalue of $N\times N$ Hermitian random matrix ensembles. Besides the analysis of the one-matrix model and their multi-critical points, as considered in this work, a possible future line of research is the analysis of the two-matrix model.

\subsection*{Acknowledgements.} The authors would like to thank Gernot Akemann for valuable discussions. The work of MA was supported by Bielefeld University. The work of SZ was partially supported by FAPERJ (grant 111.859/2012), CNPq (grant 307700/2012-7) and PUC-Rio. Further, he thanks the Mathematical Physics Group at Bielefeld University for kind hospitality and financial support during several visits which resulted in this collaboration.

\appendix

\section{Properties of the Family of Multi-critical Potentials}\label{app1}
In this appendix we give some useful results regarding the family of multi-critical potentials used above.
%\begin{lemma}
In particular, we show that the family of potentials
\bea
V_k(x)&=& \frac{2(2k+2)!}{(-1/2)_{2k+2}} \sum^{2k+1}_{l=0}\frac{(-1)^l(l+1/2)_{2k+1-l}}{((2k+1-l)!(l+1))}x^{l+1}\nn\\
&=&-4\sqrt{\pi} (2k+2)! \sum_{l=0}^{2k+1}\frac{(-1)^l x^{l+1}}{(2k+1-l)!\Gamma(l+1/2)(l+1)},\quad  k\in\mathbb{N}
\eea
has only one minimum on the real axis. Furthermore, this unique minimum is located on the positive real axis.
%\end{lemma}

%\begin{proof} 
To prove this, first note that since for any $k\in\mathbb{N}$ the potential is of even degree and the leading coefficient is positive, it follows that the potential must  have an absolute minimum on the real axis. We now proof that $V'_k(x)$ has only one zero on the real axis which then implies that this extremum is the unique real minimum. Furthermore, as a corollary of Descartes' rule it is straightforward to verify that this real minimum cannot lie on the negative real axis.

We now proceed with the proof that $V'_k(x)$ has only one zero on the real axis using Sturm's theorem. First we construct a Sturm chain:
\bea
p^0_k(x)&=&V'_k(x)\nn\\
p^1_k(x)&=&V''_k(x) \nn\\
p^2_k(x)&=&-\mathrm{rem}(p^0_k(x),p^1_k(x))\nn\\
p^3_k(x)&=&-\mathrm{rem}(p^1_k(x),p^2_k(x))\nn\\
... &=& ...\nn\\
p^m_k(x)&=&0,
\eea
where $\mathrm{rem}(p^1,p^2)$ denotes the remainder obtained from polynomial division of $p^1$ by $p^2$. One can show that in our case $m=4$ and that one has explicitly:
\bea
p^0_k(x)&=&V'_k(x)=4\sqrt{\pi} (2k+2)! \sum_{l=0}^{2k+1}\frac{(-1)^{l+1} x^{l}}{(2k+1-l)!\Gamma(l+1/2)} \nn\\
p^1_k(x)&=&V''_k(x) =4\sqrt{\pi} (2k+2)! \sum_{l=0}^{2k}\frac{(-1)^{l} (l+1) x^{l}}{(2k-l)!\Gamma(l+3/2)} \nn\\
p^2_k(x)&=& \frac{2 \sqrt{\pi} (2k+2)!}{(2k+1)^2} \sum_{l=0}^{2k-1}\frac{(-1)^{l+1}  x^{l}}{(2k-l-1)!\Gamma(l+3/2)} \nn\\
p^3_k(x)&=& -16\frac{(k+1)(2k+1)^2}{4k+1}\nn\\
p^4_k(x)&=&0
\eea
We see that the leading coefficients are 
\bea
p^0_k(x)&=& + C^0_k x^{2k+1}+...\nn\\
p^1_k(x)&=& + C^1_k x^{2k}+...\nn\\
p^2_k(x)&=& + C^2_k x^{2k-1}+...\nn\\
p^3_k(x)&=& - C^3_k\nn\\
p^4_k(x)&=&0,
\eea
where all $C$'s are positive constants. Thus Sturm's theorem implies that the number of real zeros of $V'_k(x)$ is $n=2-1=1$. This completes the proof.
%\end{proof}

%\bibliographystyle{ieeetr}
%\bibliographystyle{utphys}
%\bibliography{researchbib}

\providecommand{\href}[2]{#2}\begingroup\raggedright\endgroup

\end{document}